# Multi-Objective Optimization and Multi-Criteria Decision-Making Approach to Design Multi-Tubular Packed-Bed Membrane Reactor in Oxidative Dehydrogenation of Ethane

Seyed Reza Nabavi [a, *], Zhiyuan Wang [b], María Laura Rodríguez [c]

[a] Department of Applied Chemistry, Faculty of Chemistry, University of Mazandaran, Babolsar, Iran

[b] Department of Computer Science, DigiPen Institute of Technology Singapore, Singapore 139660, Singapore

[c] Instituto de Investigaciones en Tecnología Química (INTEQUI), UNSL-CONICET, Almirante Brown 1455, D5700HGC, San Luis, Argentina

*Corresponding author: Seyed Reza Nabavi, Email: srnabavi@umz.ac.ir

**Abstract**

Ethylene is a crucial chemical in manufacturing numerous consumer products. In recent years, the oxidative dehydrogenation of ethane (ODHE) technique has garnered significant interest as a means for ethylene production due to its high ethylene selectivity and energy efficiency, the availability of ethane as a feedstock, and catalysts that can improve the performance of the process. This work delves into determining the optimal operating conditions for the ODHE process within a multi-tubular packed-bed membrane reactor by formulating and resolving multi-objective optimization (MOO) problems. The elitist non-dominated sorting genetic algorithm II (NSGA-II) method is employed in conjunction with three different multi-criteria decision-making (MCDM) methods: (i) technique for order of preference by similarity to ideal solution (TOPSIS), (ii) preference ranking on the basis of ideal-average distance (PROBID), and (iii) simple additive weighting (SAW), to solve the MOO problems, obtain the optimal Pareto frontier, and recommend one solution for ultimate implementation. Through this approach, this work elucidates the trade-offs between conflicting objectives, deepens the comprehension of their interrelationships, and underscores the effects of decision variables on the optimization objectives. Consequently, this work provides practitioners with the insights needed for informed decision-making in the ODHE process operation.

# 1. Introduction

Ethylene, a cornerstone chemical, plays a pivotal role in the production of a wide array of consumer goods. In the contemporary petrochemical industry, ethylene production is predominantly facilitated through steam cracking of naphtha or light alkanes (Chen et al., 2023). Nonetheless, steam cracking stands as one of the most energy-intensive processes in chemical synthesis (Najari et al., 2021). Considering the undeniable challenges of the conventional steam cracking (e.g., high energy consumption, huge carbon dioxide emissions, and coke formation) in ethylene production, there has been a pressing need to explore and investigate alternative techniques. In this context, oxidative dehydrogenation of ethane (ODHE) technique has attracted much attention in recent years due to its potential to ameliorate the long-standing limitations of the conventional approach. The advantages frequently highlighted in the literature for ODHE process include its irreversibility, exothermic reactions, and production of valuable byproducts (Gao et al., 2019; Fattahi et al., 2013; Cavani et al., 2007).

The integration of membrane technologies in both conventional and contemporary reactors has gained prominence to enhance product purity through increased selectivity. The membrane used in ODHE not only serves as a filter but also as reactant distributor (Rodriguez et al., 2010), fostering a selective production of ethylene that surpasses older techniques. The introduction of a membrane interface disrupts the axial dispersion of reactants (oxygen), resulting in a significant increase in production yield. Rodriguez et al. (2010) proposed the use of inorganic porous membranes with Ni-Nb-O mixed oxide catalyst in ODHE, leading to enhanced ethylene selectivity and better heat management due to the moderate exothermicity of this reaction versus side ethane and ethylene complete oxidation reactions. Hasany et al. (2015) employed a fixed-bed catalytic membrane reactor in ODHE, observing notable enhancements in reactor performance, characterized by elevated conversions and considerable energy savings. Li & van Veen (2019) established a continuous ODHE process in the oxygen conductive membrane reactor, which increased contact time and demonstrated higher selectivity of ethylene.

Over the years, the oxidative dehydrogenation of light paraffins has been investigated from different aspects (e.g., operating conditions, membrane and catalyst types, and reactor configuration, modeling and simulation). The literature has introduced a wide array of catalysts for the process, such as metal oxides (Zhu et al., 2023; Gärtner et al., 2013; Ciambelli et al., 2000; Kennedy & Cant, 1991), alkali oxides and chlorides (Alamdari et al., 2021; Finazzi et al., 2008; Kumar et al., 2008), zeolites (Qin et al., 2023; Lin et al., 2009), and carbon nanotubes (Cao et al., 2022; Frank et al., 2010), and SBA (Deng et al., 2021; Kong et al., 2016), among others. In this regard, Ni-Nb-O mixed oxide catalyst is reported as both highly active and selective (Kong et al., 2019; Rodriguez et al., 2010; Heracleous & Lemonidou, 2006) and is utilized in the present study.

A schematic diagram of the multi-tubular packed-bed membrane reactor used in this work for ODHE is presented in Figure 1. Oxygen flows through the shell side co-currently, with the ethane stream (in the presence of small amounts of oxygen) which flows through the catalyst-filled tubes. As depicted in the inset of Figure 1, oxygen permeates the membrane from the shell side to the tube side due to a pressure difference on both sides, facilitating the oxidative dehydrogenation of ethane to ethylene on the surface of the catalyst. The multi-tubular design is chosen for better temperature control within the reactor, which significantly influences selectivity. Furthermore, by incorporating the membrane interfaces, oxygen is dosed to the reaction side, leading to lower oxygen partial pressures inside the catalyst tubes and fostering higher selectivity and better management of the heat generation (Rodriguez et al., 2010). Optimization of the process operating conditions can notably boost enhance the performance of the ODHE process.

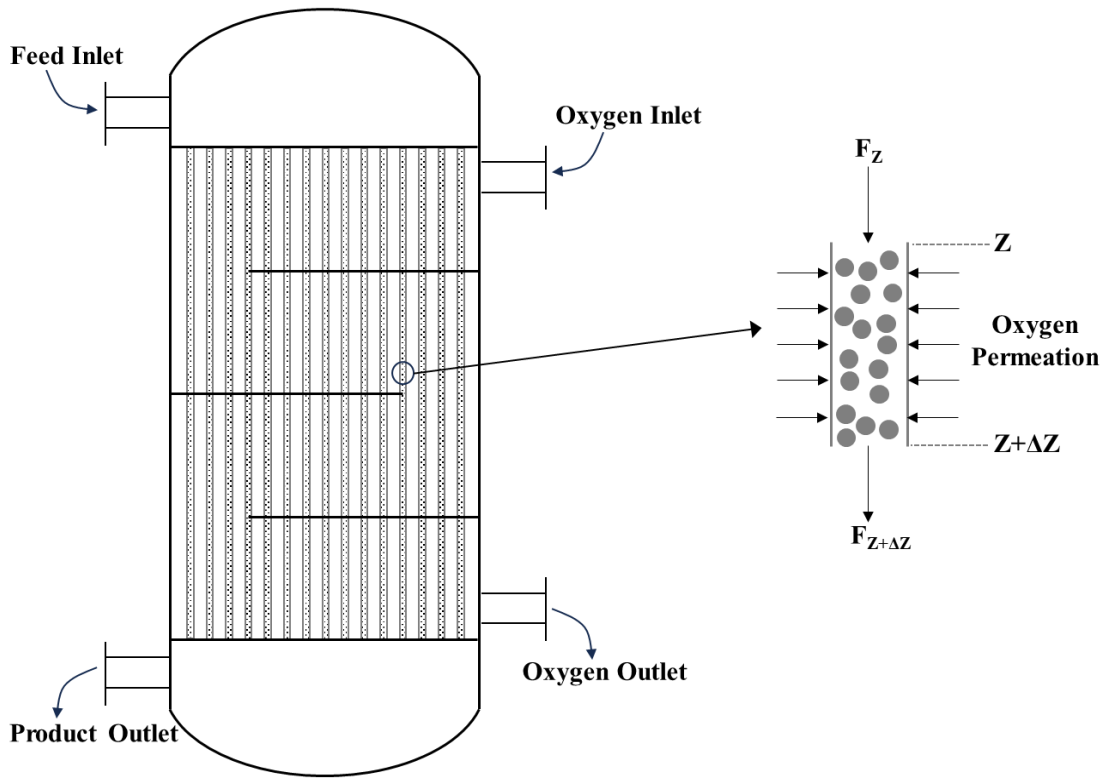

Figure 1. Scheme of multi-tubular packed-bed membrane reactor for ODHE

The primary contribution of this work is to identify the optimal operating conditions for the ODHE process within a multi-tubular packed-bed membrane reactor, considering simultaneous optimization of multiple conflicting objectives (e.g., maximizing ethylene production and minimizing carbon dioxide output).

To address such intricate optimization challenges, multi-objective optimization (MOO) methods are typically employed. Reviews of MOO applications in chemical engineering indicate that the elitist non-dominated sorting genetic algorithm II (NSGA-II), proposed by Deb et al. (2002), is a prevalent choice. For example, Yuan et al. (2014) employed NSGA-II to concurrently minimize energy consumption and the required membrane surface area in nitrogen-selective membrane for post-combustion carbon dioxide capture. He & You (2015) utilized NSGA-II in devising a more economical and environmentally friendly process for producing chemicals from shale gas and bioethanol. Nabavi (2016) used NSGA-II to optimize the preparation conditions of asymmetric polyetherimide membrane for prevaporation of isopropanol considering two objectives: simultaneous maximization of membrane flux and separation factor. Xiao et al. (2018) applied NSGA-II to optimize the methyl tert-butyl ether reactive distillation process, aiming to maximize the conversion of isobutene and minimize total annual cost. Kherdekar et al. (2021) used NSGA-II in designing a load-flexible fixed-bed reactor to facilitate the partial water-gas shift reaction within a coal-to-methanol plant. Wang et al. (2022) employed NSGA-II in two chemical processes: a supercritical water gasification process aiming for $H_2$-rich syngas with reduced greenhouse gas emissions, and a combustion process targeting for higher energy output and lower environmental pollution. NSGA-II was recently integrated into a model predictive control framework to address multiple objectives for a continuous stirred tank reactor (Wang, Tan, et al., 2023).

When accounting for multiple conflicting objectives, MOO methods, such as NSGA-II, yield a set of non-dominated solutions that are equally optimal from the standpoint of the specified objectives (Nabavi, Jafari, et al., 2023). These solutions constitute an optimal Pareto frontier, where improving one objective inevitably compromises another. Nevertheless, when it comes to practical implementation, only one solution from the optimal Pareto frontier is required. This is often guided by MCDM methods, which rank and recommend a single solution from the optimal Pareto frontier. Rangaiah et al. (2020) highlighted that most MOO applications in chemical engineering have been centered around obtaining the optimal Pareto frontier, with MCDM being somewhat overlooked. They suggested that this could be attributed to the perception that decision and selection from the optimal Pareto frontier can be made sufficiently based upon engineering experience and individual preference. In our prior publications (Nabavi, Wang, et al., 2023; Wang, Nabavi, et al., 2023), we undertook a systematic investigation on a number of MCDM methods. In this work, three MCDM methods, namely, technique for order of preference by similarity to ideal solution (TOPSIS) proposed by Hwang & Yoon (1981), preference ranking on the basis of ideal-average distance (PROBID) created by Wang et al. (2021), and simple additive weighting (SAW) introduced in MacCrimmon (1968) and Fishburn (1967), are employed to select one solution from the optimal Pareto frontier for ultimate practical implementation.

The article unfolds as follows: Section 2 provides the mathematical modeling of the ODHE process. Section 3 states the MOO formulation and MCDM methods. Section 4 presents sensitivity analyses of decision variables on the optimization objective functions and showcases the optimization results for three distinct scenarios. The considered scenarios are: first, the co-maximization of ethane conversion rate and ethylene selectivity; second, the maximization of yearly ethylene production rate and minimization of yearly carbon dioxide production rate; and third, the simultaneous maximization of ethane conversion and ethylene production rate, as well as minimization of carbon dioxide production rate. Finally, concluding remarks along with suggestions for prospective research are consolidated in Section 5.

## 2. Mathematical Modeling of ODHE in Multi-Tubular Packed-Bed Membrane Reactor

The system of reactions for ODHE process is presented in equations (1) to (3). Clearly, the complete oxidation of ethane and the oxidation of ethylene, as represented by equations (2) and (3), are undesirable reactions. These side reactions not only reduce the ethylene yield but also leads to additional $CO_2$ production and the undesirable heating of the reaction medium. Elevated temperatures can jeopardize the catalyst and reduce selectivity. Precise control of oxygen dosing in the reaction medium is pivotal in mitigating these side reactions. Optimal dosing, at 0.5 moles $O_2$ per 1 mole $C_2H_6$, can steer the process towards preferred mechanisms. In the proposed reactor design, the Raschig ring catalysts with an average active layer thickness of 180 micrometers are considered.

$$C_2H_6 + 0.5O_2 \rightarrow C_2H_4 + H_2O \quad \Delta H = -105 \text{ kJ/mol} \quad (1)$$

$$C_2H_6 + 3.5O_2 \rightarrow 2CO_2 + 3H_2O \quad \Delta H = -1428 \text{ kJ/mol} \quad (2)$$

$$C_2H_4 + 3O_2 \rightarrow 2CO_2 + 2H_2O \quad \Delta H = -1323 \text{ kJ/mol} \quad (3)$$

The ODHE reactor comprises 1000 membrane tubes housed within a shell, as shown in Figure 1. The ethane stream (on the tube side) flows co-currently with oxygen (on the shell side), minimizing the likelihood of hot-spot formation (Rodriguez et al., 2010). The assumptions include:

- The ODHE operates under non-isothermal (both shell and tube) and steady state conditions.
- The reactor shell is assumed to be properly isolated with negligible heat losses to the environment.
- Given the small tube diameter and elevated internal flow rate, a plug flow regime is assumed (flat radial distributions of concentration and temperature).
- Axial mass and energy dispersions and external transport limitations are assumed to be negligible owing to the high gas velocity in both the shell and tubes sides.
- Internal mass- and energy-transport limitations are also neglected because a thin washcoat over the catalyst particles (egg-shell type) is chosen
- The ideal gas law is employed for predict the partial pressure of the gaseous compounds.
- A cocurrent flow configuration between the process gas and the oxygen in the shell is selected.
- Coke generation and catalyst deactivation are negligible.

The Power Law type kinetic model proposed by Heracleous & Lemonidou (2006) for the $Ni_{0.85}Nb_{0.15}O$ catalyst is used for the simulations:

$$R_i = k_i\, P_{HC}^{a_i}\, P_{O_2}^{b_i} \quad (4)$$

$$k_i = k_{0i} \exp\left[\frac{E_{ai}}{R}\left(\frac{1}{T_0} - \frac{1}{T}\right)\right] \quad (5)$$

Table 1 reports the corresponding parameters for $R_i$ ($T_0$ = 543 K, $P_j$ in Pa).

| Table 1. Kinetic parameters for ODHE reactions (Heracleous & Lemonidou, 2006)$a$ | $b$ | $E_a$ (kJ/mol) | $k_0\,(mol\, kg_{cat}^{-1}\, s^{-1}\, Pa^{-(a+b)})$ |
|---|---|---|---|
| 0.52 | 0.213 | 96.18 | $4.177 \times 10^{-10}$ |
| 0.547 | 0.829 | 76.21 | $1.272 \times 10^{-13}$ |
| 0.475 | 0.319 | 98.42 | $9.367 \times 10^{-11}$ |

Reaction (*i*)   $k_{0,i}$ [kmol $kg_{cat}^{-1}$ s$^{-1}$ Pa$^{-(a+b)}$]   $E_{ai}$ [kJ/mol]   $a_i$   $b_i$

| 1 | 4.177 x 10⁻¹⁰ | 96.18 | 0.520 | 0.213 |
| 2 | 1.272 x 10⁻¹³ | 76.21 | 0.547 | 0.829 |
| 3 | 9.367 x 10⁻¹¹ | 98.42 | 0.475 | 0.319 |

A semi homogenous two-dimensional model is applied in steady state conditions. The mass, heat, and momentum transfer equations are built in the selected element to simulate ODHE process (presented in the Supporting Information), resulting in a set of initial value ordinary differential equations (Rodriguez et al., 2010). Conversion rate of ethane ($X_{C_2H_6}$) and ethylene selectivity ($S_G$) are defined as follows, respectively:

$$X_{C_2H_6} = \frac{F^0_{C_2H_6} - F_{C_2H_6}}{F^0_{C_2H_6}} \tag{6}$$

$$S_G = \frac{F_{C_2H_4}}{F^0_{C_2H_6} - F_{C_2H_6}} \tag{7}$$

Where $F^0_{C_2H_6}$ is the input flowrate of ethane; $F_{C_2H_6}$ is the output flowrate of ethane and $F_{C_2H_4}$ is the output flowrate of ethylene. Process yield is calculated by multiplying equations (6) and (7).

The initial operating conditions of the equations governing the ODHE process and the initial geometrical parameters are presented in Table 2. The differential equations are integrated in the spatial coordinate by means of a Gear-type integration routine.

Table 2. Geometrical parameters and operating conditions.

| Parameter | Value | Parameter | Value |
|---|---|---|---|
| $d$, m or $d$ (m) | 0.0266 | $F_{S0}$ | 3010 |
| $ds$, m | 3.93 | $y_{o2,0}$ | 0 |
| $L$, m | 4 | $T_0$ | 380 |
| $n_T$ | 10000 | $T_{S0}$, °C | 25 |
| $\rho_B$, $kg_{cat}/m^3 R$ | 200 | $P_0$, atm | 5 |
| $d_p$, | 0.0045 | $P_{S0}$ | 5.4 |
| $F_0$, | 3000 | $n_b$ | 3 |
| $\varepsilon$ | 0.48 | | |

## 3. MOO Formulation and MCDM Methods

### 3.1. MOO Formulation

MOO is a process in which by changing the input decision/operating variables, multiple goals (i.e., objective functions) are simultaneously achieved. A basic formulation for MOO problems is presented below:

$$\text{Minimize } F(\vec{X}) = [f_1(\vec{X}), f_2(\vec{X}), \dots, f_k(\vec{X})]^T \tag{8a}$$

Subject to:

$$\vec{g}(\vec{X}) \leq 0 \tag{8b}$$

$$\vec{h}(\vec{X}) = 0 \tag{8c}$$

$$\vec{X} \in R^n, \vec{f}(\vec{X}) \in R^k, \vec{g}(\vec{X}) \in R^m, \text{ and } \vec{h}(\vec{X}) \in R^q$$

Where, $k$, $n$, $m$ and $q$ are respectively the number of objective functions, decision variables, inequalities and equalities constraints. A vector of decision variable and a vector of objective functions are defined as $\vec{X} \in R^n$ and $\vec{f}(\vec{X}) \in R^k$, respectively. Unlike to single objective optimization (SOO) problems, solving an MOO problem leads to a set of non-dominated solutions that satisfy all objective functions simultaneously.

In this work, six operating/decision variables, namely, the oxygen molar fraction at tube side ($y_{O2,0}$), total inlet molar flow rate at tube side ($F_0$) in kmol/h, inlet temperature at the tube side ($T_0$) in °C, inlet pressure at the shell side ($P_{S0}$) in atm, inlet temperature at the shell side ($T_{S0}$) in °C, and total inlet molar flow rate at shell side ($F_{S0}$) in kmol/h are chosen to find their optimal operating conditions. Besides, the lower bound and upper bound of each decision variable are presented in Table 3.

Table 3. The lower and upper bounds of input decision variables for ODHE process

| Decision variable | Lower bound | Upper bound |
|---|---|---|
| $y_{O2,0}$ | 0 | 0.05 |
| $F_0$ (kmol/h) | 1000 | 4500 |
| $T_0$ (°C) | 360 | 420 |
| $P_{S0}$ (atm) | 5.15 | 6 |
| $T_{S0}$ (°C) | 25 | 200 |
| $F_{S0}$ (kmol/h) | 3000 | 5000 |

Three MOO scenarios are investigated in this work: namely, the co-maximization of ethane conversion rate and ethylene selectivity; the maximization of yearly ethylene production rate and minimization of yearly carbon dioxide production rate; and the simultaneous maximization of ethane conversion and ethylene production rate and minimization of carbon dioxide production rate. NSGA-II is applied to solve these MOO problems to obtain the optimal Pareto frontiers. The seven steps of NSGA-II procedure are illustrated in Figure 2. A solution population of 100 and a maximum number of generations of 150 are set. Probabilities for crossover and mutation operators are assigned as 0.7 and 0.4, respectively. These parameter settings for NSGA-II are chosen based on our preliminary tests to strike a balance between computational efficiency and the reliability of the Pareto optimal curve. Following these initial settings, the MOO is executed in MATLAB.

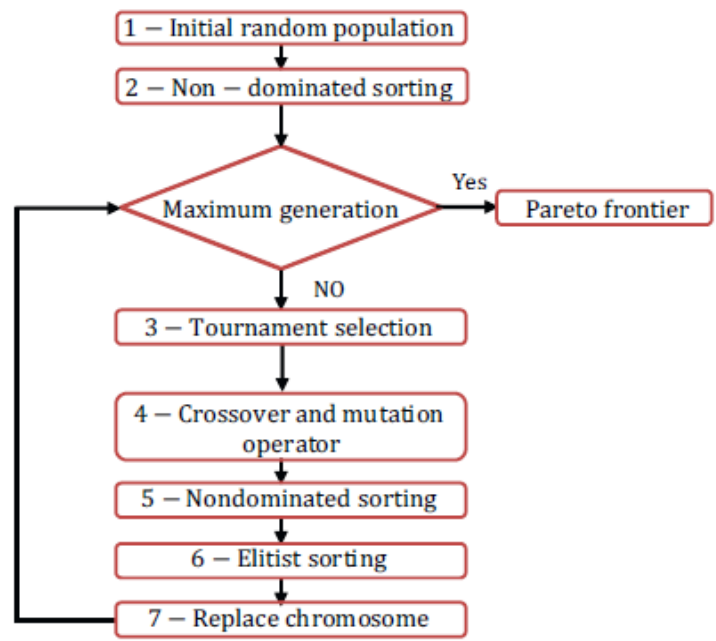

Figure 2. NSGA-II procedure

### 3.2. MCDM Methods

#### 3.2.1. TOPSIS Method

TOPSIS was first proposed by Hwang & Yoon (1981). The core principle of the TOPSIS method is to find the solution that is closest to the positive-ideal solution and farthest from the negative-ideal solution (Wang & Rangaiah, 2017). Its procedure can be divided into following five steps:

**Step 1.** Construct normalized objective matrix ($F_{ij}$) with $m$ rows (solutions) and $n$ columns (objectives) by applying vector normalization.

$$F_{ij} = \frac{f_{ij}}{\sqrt{\sum_{k=1}^{m} f_{kj}^2}} \qquad (9)$$

**Step 2.** Construct weighted normalized objective matrix ($v_{ij}$) by multiplying each column with its weight $w_j$:

$$v_{ij} = F_{ij} \times w_j \qquad (10)$$

**Step 3.** Determine the positive-ideal solution $A^+$ and negative-ideal solution $A^-$ as follows. Firstly, find the best value of each objective. For maximization objective, the best value is the largest value within the column of the objective matrix. Otherwise, for minimization objective, the best value is the smallest value in the column. Mathematically, these are given by:

$$A^+ = \{(Max_i(v_{ij})|j \in J), (Min_i(v_{ij})|j \in J') \mid i \in 1,2,\ldots,m\}$$
$$= \{v_1^+, v_2^+, v_3^+, \ldots, v_j^+, \ldots, v_n^+\} \qquad (11)$$

Here, $J$ is the set of maximization objectives and $J'$ is the set of minimization objectives, from the total set of $\{1, 2, 3, 4, \ldots, n\}$.

Next, find the worst value of each objective, which is the smallest and largest value within the column of the objective matrix, respectively for maximization and minimization objective. These values constitute the negative-ideal solution given by:

$$A^- = \{(Min_i(v_{ij})|j \in J), (Max_i(v_{ij})|j \in J') \mid i \in 1,2,\ldots,m\}$$
$$= \{v_1^-, v_2^-, v_3^-, \ldots, v_j^-, \ldots, v_n^-\} \qquad (12)$$

**Step 4.** Calculate the Euclidean distance between each solution and positive-ideal/negative-ideal solution:

$$\text{Distance to positive ideal}, S_{i+} = \sqrt{\sum_{j=1}^{n}(v_{ij} - v_j^+)^2} \quad i = 1,2,3,\ldots,m \qquad (13)$$

$$\text{Distance to negative ideal}, S_{i-} = \sqrt{\sum_{j=1}^{n}(v_{ij} - v_j^-)^2} \quad i = 1,2,3,\ldots,m \qquad (14)$$

**Step 5.** Calculate the closeness of each optimal solution:

$$C_i = \frac{S_{i-}}{S_{i-} + S_{i+}} \qquad (15)$$

When $S_{i-} = 0$, $C_i = 0$ and solution $i$ is the closest to negative-ideal solution. When $S_{i+} = 0$, $C_i = 1$ and $i$ is the closest to the positive-ideal solution. The solution having the largest $C_i$ value is the recommended solution.

### 3.2.2. PROBID Method

Introduced by Wang et al. (2021), the PROBID method stands out due to its comprehensive consideration of the mean solution and all the different tiers of ideal solutions when ranking non-dominated solutions (Wang, Baydaş, et al., 2023). These ideal solutions range from the most positive ideal solution (PIS) to the 2$^{nd}$ PIS, 3$^{rd}$ PIS, 4$^{th}$ PIS, and so on, culminating in the last PIS, also known as the most negative ideal solution (NIS). The performance score is then determined by computing and amalgamating distances between each non-dominated solution and these ideal solutions, along with the distance from the mean solution. The six steps of PROBID are detailed as follows:

**Step 1.** Construct the normalized objective matrix ($v_{ij}$) with *m* rows (solutions) and *n* columns (objectives) by applying vector normalization.

$$F_{ij} = \frac{f_{ij}}{\sqrt{\sum_{k=1}^{m} f_{kj}^2}} \quad i \in \{1,2,\dots,m\}; j \in \{1,2,\dots,n\} \tag{16}$$

**Step 2.** Construct the weighted normalized objective matrix by multiplying each column with its weight, $w_j$:

$$v_{ij} = F_{ij} \times w_j \quad i \in \{1,2,\dots,m\}; j \in \{1,2,\dots,n\} \tag{17}$$

**Step 3.** Determine the most positive ideal solution PIS ($A_{(1)}$), 2$^{nd}$ PIS ($A_{(2)}$), 3$^{rd}$ PIS ($A_{(3)}$), ..., and $m^{th}$ PIS ($A_{(m)}$) (i.e., the most NIS).

$$A_{(k)} = \{ (Large(v_j, k)|j \in J),\ (Small(v_j, k)|j \in J') \} =$$
$$\{v_{(k)1}, v_{(k)2}, v_{(k)3}, \dots, v_{(k)j}, \dots, v_{(k)n}\} \tag{18}$$

Here, $k \in \{1,2,\dots,m\}$, $J$ = set of maximization objectives from $\{1, 2, 3, 4, \dots, n\}$, $J'$ = set of minimization objectives from $\{1, 2, 3, 4, \dots, n\}$, $Large(v_j, k)$ means the $k^{th}$ largest value in the $j^{th}$ weighted normalized objective column (i.e., $v_j$) and $Small(v_j, k)$ means the $k^{th}$ smallest value in the $j^{th}$ weighted normalized objective column (i.e., $v_j$).

$$\bar{v}_j = \frac{\sum_{k=1}^{m} v_{(k)j}}{m} \quad \text{for } j \in \{1,2,\dots,n\} \tag{19}$$

The average solution is then:

$$\bar{A} = \{\bar{v}_1, \bar{v}_2, \bar{v}_3, \dots, \bar{v}_j, \dots, \bar{v}_n\} \tag{20}$$

**Step 4.** Calculate the Euclidean distance of each solution to each of the *m* ideal solutions as well as to the average solution. Distance to ideal solutions is found as:

$$S_{i(k)} = \sqrt{\sum_{j=1}^{n} (v_{ij} - v_{(k)j})^2} \quad i \in \{1,2,\dots,m\}; k \in \{1,2,\dots,m\} \tag{21}$$

Next, the distance to average solution is found as:

$$S_{i(avg)} = \sqrt{\sum_{j=1}^{n}(v_{ij} - \bar{v}_j)^2} \quad i \in \{1,2,\dots,m\} \tag{22}$$

**Step 5.** Determine the overall positive-ideal distance, which is essentially the weighted sum distance of one solution to the first half of ideal solutions.

$$S_{i(pos-ideal)} = \begin{cases} \sum_{k=1}^{\frac{m+1}{2}} \frac{1}{k} S_{i(k)} \ i \in \{1,2,\dots,m\} \ when\ m\ is\ an\ odd\ number \\ \sum_{k=1}^{\frac{m}{2}} \frac{1}{k} S_{i(k)} \ i \in \{1,2,\dots,m\} \ when\ m\ is\ an\ even\ number \end{cases} \tag{23}$$

Here, weight is decreasing with the ideal solution number (i.e., $k$ = 1, 2, 3, etc.). Similarly, determine the overall negative-ideal distance, which is essentially the weighted sum distance of one solution to the second half of ideal solutions.

$$S_{i(neg-ideal)} = \begin{cases} \sum_{k=\frac{m+1}{2}}^{m} \frac{1}{m-k+1} S_{i(k)} \ i \in \{1,2,\dots,m\} \ when\ m\ is\ an\ odd\ number \\ \sum_{k=\frac{m}{2}+1}^{m} \frac{1}{m-k+1} S_{i(k)} \ i \in \{1,2,\dots,m\} \ when\ m\ is\ an\ even\ number \end{cases} \tag{24}$$

Here, weight is increasing with the ideal solution number (i.e., $k$ increasing to $m$).

**Step 6.** Calculate the *pos-ideal/neg-ideal* ratio ($R_i$) and then performance score ($P_i$) of each solution as follows:

$$R_i = \frac{S_{i(pos-ideal)}}{S_{i(neg-ideal)}} \quad i \in \{1,2,\dots,m\} \tag{25}$$

$$P_i = \frac{1}{1+R_i^2} + S_{i(avg)} \quad i \in \{1,2,\dots,m\} \tag{26}$$

The solution having the largest $P_i$ value is the recommended solution.

### 3.2.3. SAW Method

The SAW method is probably the simplest MCDM method (Nabavi, Wang, et al., 2023). It was originally proposed by Fishburn (1967) and MacCrimmon (1968). The fundamental concept of the SAW method is to compute a performance score for each solution by summing the products of its objective (i.e., criterion) values and their assigned weights. The stepwise procedure of SAW is given as follows:

**Step 1.** Construct the normalized objective matrix of $m$ rows (solutions) and $n$ columns (objectives), by applying:

$$F_{ij} = \frac{f_{ij}}{f_j^+} \ for\ a\ maximization\ criterion, where\ f_j^+ = Max_{i \in m} f_{ij} \tag{24}$$

$$F_{ij} = \frac{f_j^-}{f_{ij}} \ for\ a\ minimization\ criterion, where\ f_j^- = Min_{i \in m} f_{ij} \tag{25}$$

**Step 2.** Construct the weighted normalized objective matrix by applying:

$$v_{ij} = F_{ij} \times w_j \tag{26}$$

**Step 3.** Find the score of each optimal solution, i.e., sum of weighted value by:

$$A_i = \sum_{j=1}^{n} v_{ij} \tag{27}$$

The solution having the largest $A_i$ value is the recommended solution.

## 4. Results and Discussion

Solving the ordinary differential equations reported in Rodriguez et al. (2010) by means of Gear algorithm under the operating conditions listed in Table S3, ODHE process in a packed bed membrane reactor is simulated. Temperature ($T$ and $T_s$), inlet oxygen mole fraction ($y_{O2,0}$), ethylene selectivity ($S_G$), ethane conversion rate, and ethylene production rate profiles along the reactor are presented in Figure S1. Figure S1a illustrates the axial temperature profiles of both the shell and tubes along the length of the reactor. The red solid curve represents the mean radial temperature on the tube side. The main $O_2$ feed is permeated through the membrane and no $O_2$ is introduced at the reactor inlet ($y_{O2,0}=0$). A constant trans-membrane pressure drops of 0.4 atm is selected.

In the first section of the reactor length (~ 1 m), the temperature profile shows a minimum followed by a continuous and smooth increase up to the reactor outlet. Due to the low heat capacity of the oxygen flowing through the shell side, the temperature profile presents a high temperature difference between the shell inlet and outlet and tends to overlap with that of the reactor side beyond 1.5m. The slight and less pronounced change in the slope of the temperature profile around 3 m is due to increased selectivity toward the desired, less exothermic reaction. Increased selectivity (see Figure S1c) is associated with the axial position where full consumption of the accumulated $O_2$ inside the tubes is recorded as shown Figure S1b. Beyond this position the permeated $O_2$ is mainly consumed by the desired reaction (reaction 1), and a sustained ethane conversion and ethylene production is registered (see Figures S1d and S1e).

### 4.1. Sensitivity Analyses

The impact of each decision variable on the optimization objectives —ethane conversion rate ($X_{C_2H_6}$), ethylene selectivity ($S_G$), and production rates of ethylene and carbon dioxide— is investigated in this subsection. The other operating conditions and geometrical parameters remains invariant (see Table 2). As illustrated in Figure 3, elevating the tube-side inlet temperature ($T_0$) increases the reaction rates, thereby increasing the oxygen consumption rate. Consequently, as oxygen is permeated through the membrane is consumed. The low oxygen content inside the tubes favors the desirable reaction (equation (1)). The increase in the tube-side inlet temperature boosts ethylene selectivity (see Figure 3b), increasing ethane conversion rate (see Figure 3a), , and ethylene production rate (see Figure 3c), while decreasing carbon dioxide emissions (see Figure 3d) due to reaction 1 does not generate $CO_2$. On the other hand, increasing the inlet temperature on the shell side ($T_{S0}$) between 25 and 200 °C reduces both the heat transfer rate between the shell and tubes and the cold-shot effect, with together helps to increase the temperature level inside the tubes, resulting similar tendences but much lightereffects on the optimization objectives compared to the impact of increasing $T_0$ from 360 to 420 °C.

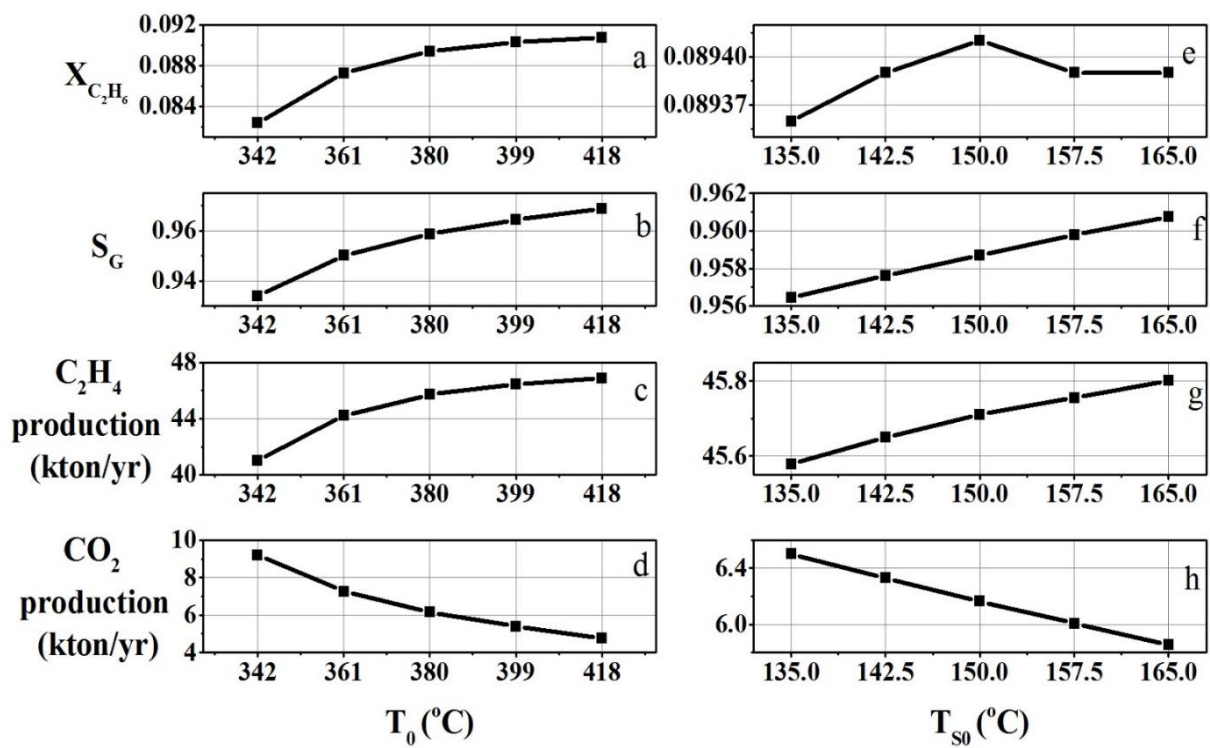

Figure 3. Impact of decision variables $T_0$ and $T_{S0}$ on the optimization objectives

Increasing the oxygen feed content ($y_{O2,0}$) from 0 to 0.05 has two opposite effects. On the one hand, when no oxygen is fed at $z = 0$, reaction rates are lower and $O_2$ consumption is slow. The rate of permeation across the membrane exceeds the rate of oxygen consumption in the tube side. Then, the gradual increase of the partial pressure of $O_2$ leads to higher reaction rates and the oxygen content inside the tube decreases again near the reactor outlet. This accumulation phenomenon leads to ethylene selectivity losses, as demonstrated by Rodriguez et al., 2010. On the other hand, a higher oxygen level at the reactor inlet tends to favor side reactions at this position, where there is a slight decrease in ethylene selectivity, although once the $O_2$ fed by the reactor inlet is consumed, the consequent increase in the thermal level inside the tubes improves the reaction rates, so that the permeated $O_2$ is immediately consumed mainly by reaction 1 and the aforementioned $O_2$ accumulation phenomenon does not occur. As an overall result, the slight increase of the inlet $O_2$ generates an improvement in ethylene production (as long as it is kept at low values as recommended by Rodriguez et al., 2010) suggesting that the higher reaction rates reached at the reactor inlet improve the thermal level inside the tubes and ethane conversion is accelerated (see Figure 4a), and despite the predicted loss of selectivity in the first section of the tubes, globally this loss is negligible (see Figure 4b) and the yield towards ethylene tends to increase (see Figure 4c). In Figure 4d the slight increase in the presence of $CO_2$ confirms the loss of selectivity towards the desired reaction.

The effect of changing the shell side stream pressure ($P_{S0}$) is illustrated on

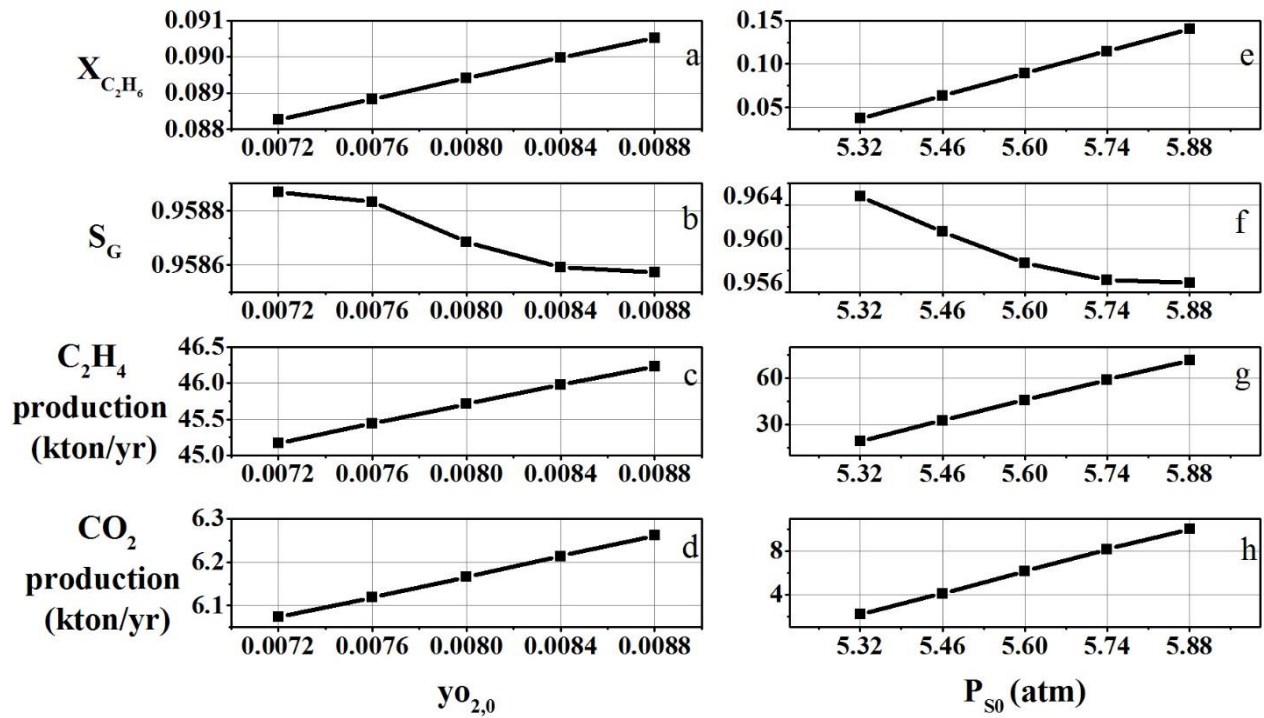

Figure 4e to Figure 4h. Increasing the $P_{S0}$ value promotes the oxygen permeation to the tube side. The production rate is determined by the permeated oxygen. In fact, the outlet production rate increases more than 100% when the trans-membrane pressure drop is increased from 0.32 to 0.88 atm (see Figure 4g). It is worth mentioning that the outlet global selectivity does not suffer a notorious lost for this threefold variation in $P_{S0}$ as no oxygen accumulation occurs ($0.956 < S_G < 0.965$). Thus, increasing $P_{S0}$ leads to similar trends as increasing to increasing the oxygen fraction in the tube side feed stream, albeit with a more pronounced effect.

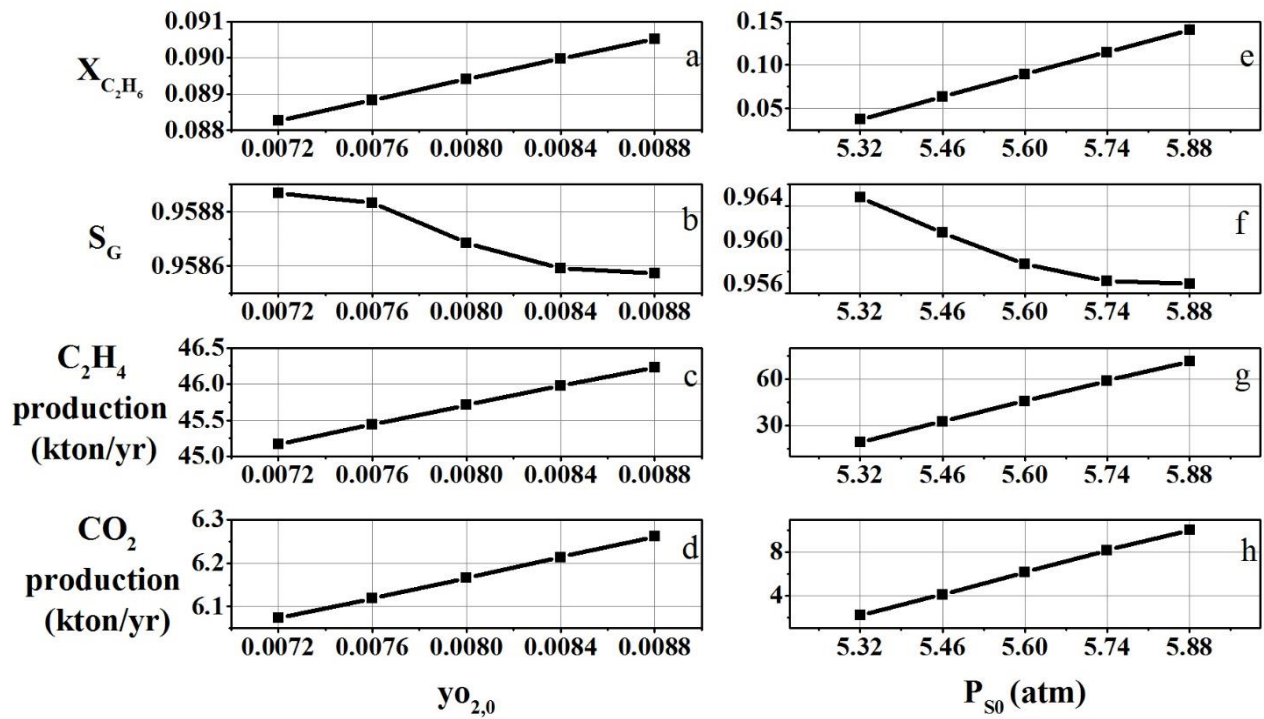

Figure 4. Impact of decision variables $y_{O2,0}$ and $P_{S0}$ on the optimization objectives

Figure 5 illustrates the effects of varying the molar flow rates of both the shell and tube side feeds on the optimization objectives. Increasing the tube-side molar flow rate (i.e., ethane molar flow rate) diminishes the residence time and reduces the partial pressure of the permeated oxygen in the tube side. As a consequence, the ethane conversion rate declines, ethylene selectivity increases, and the production rates of both ethylene and carbon dioxide rise.
It could be mentioned that the double cooling effect (by convection-conduction between the tubes and the shell sides and by cold-shot) is attenuated as F0 increases.

In contrast, elevating the shell side molar flowrate (i.e., oxygen molar flowrate) increases the heat transfer from the tube to the shell side. This more effectively cools the catalytic bed, resulting in a notable reduction in reaction rate and conversion (almost halve them), and subsequently lowers the production rates of both ethylene and carbon dioxide. Meanwhile, the slower reaction rates and higher oxygen molar flowrate result in a greater oxygen presence within the reactor, which decreases the ethylene selectivity.

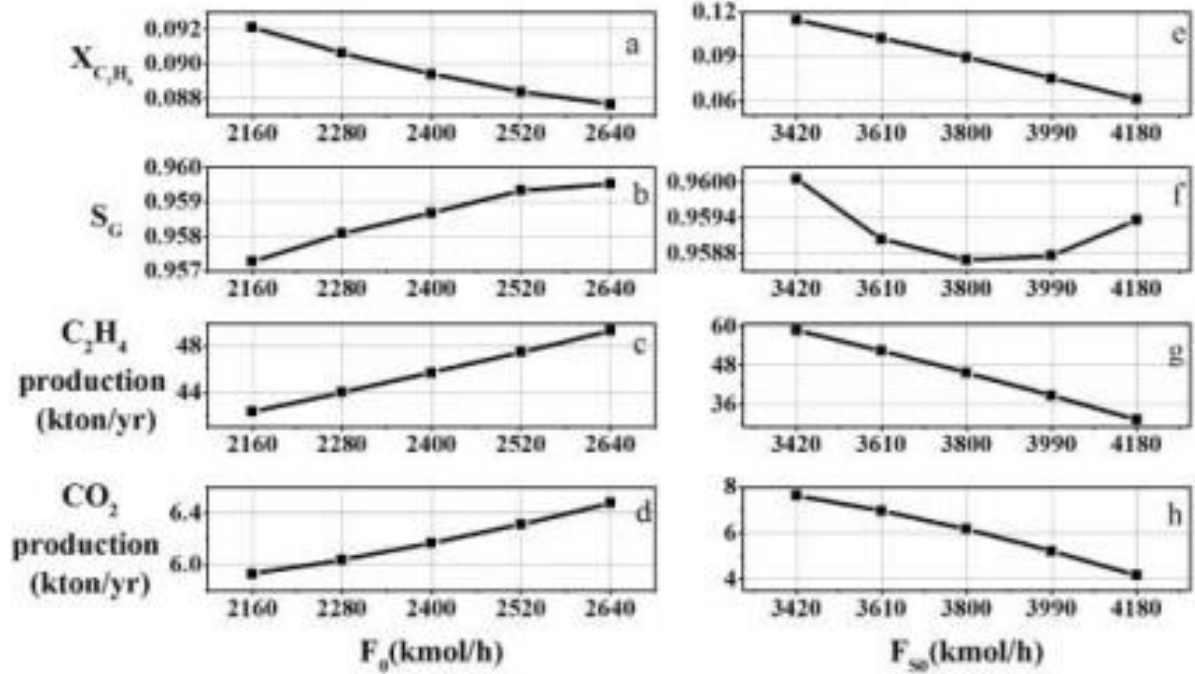

Figure 5. Impact of decision variables $F_0$ and $F_{S0}$ on the optimization objectives

**4.2. Co-maximization of Ethane Conversion Rate and Ethylene Selectivity**

    Figure 6 presents the optimal Pareto frontiers aiming for co-maximization of ethane conversion rate and ethylene selectivity. It includes the initial population as well as the optimal Pareto frontiers at the 25th, 100th, and 150th generations. Due to the imposition of a temperature constraint and the divergence observed in the model used, the initial population is 84% fewer than the intended number of 100 solutions. However, as the generations progress, the number of solutions added to the optimal Pareto frontier increases, with population counts reaching 55 at the 25th generation and 100 at the 100th generation. Additionally, as the generations increase, the span of the optimal Pareto frontier broadens. The highest values for ethane conversion rate and ethylene selectivity are 0.1903 and 0.991, respectively.

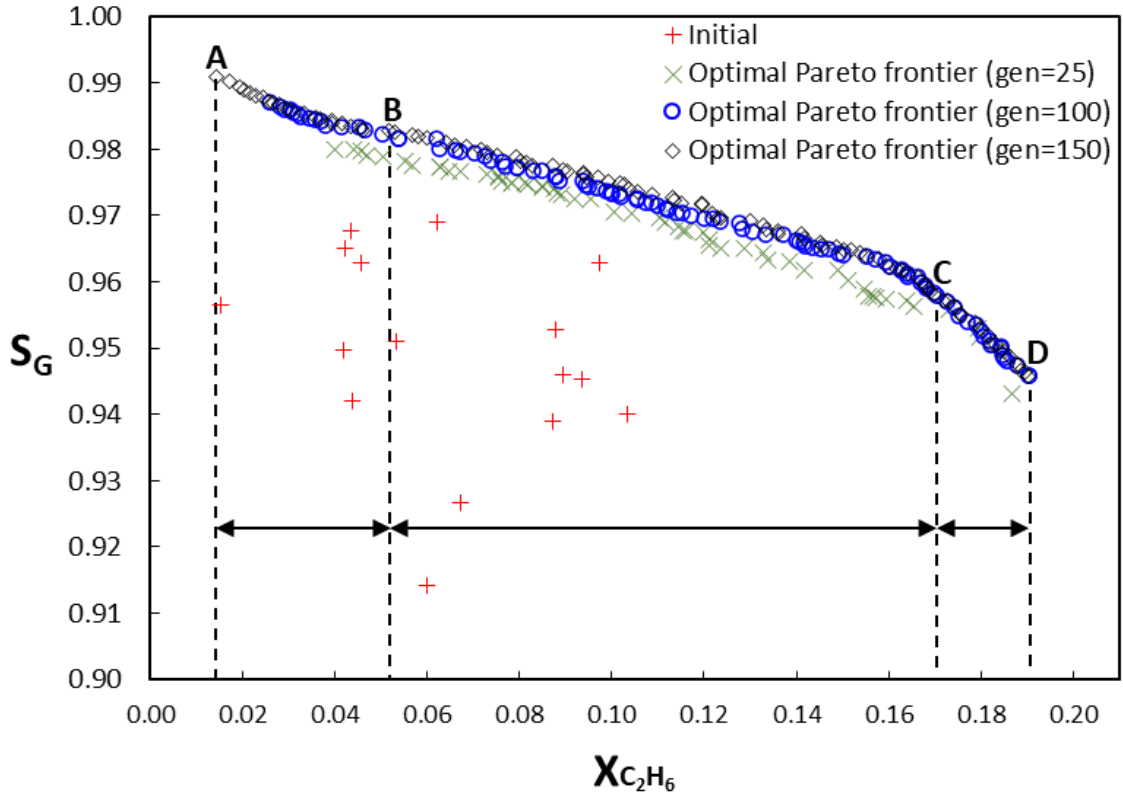

Figure 6. Optimal Pareto frontiers for co-maximization of ethane conversion rate and ethylene selectivity

Moreover, as previously discussed, ethylene selectivity is significantly influenced by the oxygen distribution and increases when oxygen partial pressure in the catalytic bed is low.. Figure 6 shows three distinct ethane conversion regions, that can be identified as Region A to B (0.014 to 0.052), Region B to C (0.052-0.168), and Region C to D (0.168 to 0.1903). Within each region, the NSGA-II optimization algorithm attempts to maximize both the ethane conversion rate and ethylene selectivity. The interaction between these two objectives prompts the algorithm to adjust its optimization path to achieve the most favorable values for both ethane conversion and ethylene selectivity. Table 4 presents the objective values for the decision variables in solutions A, B, C, and D. As can be seen, when moving from region A to D, ethane conversion increases, while ethylene selectivity decreases.

Table 4. Objective values for the decision variables of in solutions A, B, C, and D for co-maximization of ethane conversion and ethylene selectivity

|  | A | B | C | D |
|---|---|---|---|---|
| $X_{C_2H_6}$ | 0.014 | 0.052 | 0.168 | 0.1903 |
| $S_G$ | 0.991 | 0.983 | 0.959 | 0.946 |
| $Y_{O2,0}$ | 0.00012 | 0.0012 | 0.0095 | 0.0189 |
| $F_0$ (kmol/h) | 1162 | 2399 | 1000 | 1000 |
| $T_0$ (°C) | 420 | 420 | 404.2 | 392.5 |
| $P_{S0}$ (atm) | 5.291 | 5.15 | 5.577 | 5.586 |
| $T_{S0}$ (°C) | 200 | 200 | 197.4 | 144.2 |
| $F_{S0}$ (kmol/h) | 3671 | 3038 | 3581 | 3537 |

Next, to select one solution from the final optimal Pareto frontier for practical implementation, the three MCDM methods mentioned above, namely, TOPSIS, PROBID, and SAW, are applied to rank the solutions on the final optimal Pareto frontier and subsequently recommend one solution. In the current case of co-maximization of ethane conversion and ethylene selectivity, equal weighting is assigned to both objectives variables in the MCDM process (i.e., 0.5 to each).

It is observed that TOPSIS and PROBID recommend the same solution (shown as red triangle ▲ in Figure 7). In contrast, the solution recommended by SAW exhibits a slight difference (shown as blue square ■ in Figure 7). Table 5 lists the values of the objective and decision variables, as well as the solutions recommended by TOPSIS, PROBID, and SAW. The last method suggests an ethane conversion rate of 0.1903, slightly higher than 0.1896 predicted by TOPSIS and PROBID. As expected, this small increase in ethane conversion is accompanied by a decrease in ethylene selectivity: 0.9458 according to SAW, versus the the slightly higher 0.9464 by TOPSIS and PROBID.

Additionally, with respect to decision variables, both the solution selected by SAW and that of TOPSIS and PROBID agree on $F_0$ and $P_{S0}$, but show minor discrepancies for the remaining four decision variables ($y_{O2,0}$, $T_0$, $T_{S0}$, and $F_{S0}$). Overall, based on majority vote and similarity to that chosen by SAW, the solution selected by TOPSIS and PROBID, accompanied by the corresponding values of the decision variables presented in Table 5, stands as the preferred recommendation for implementation.

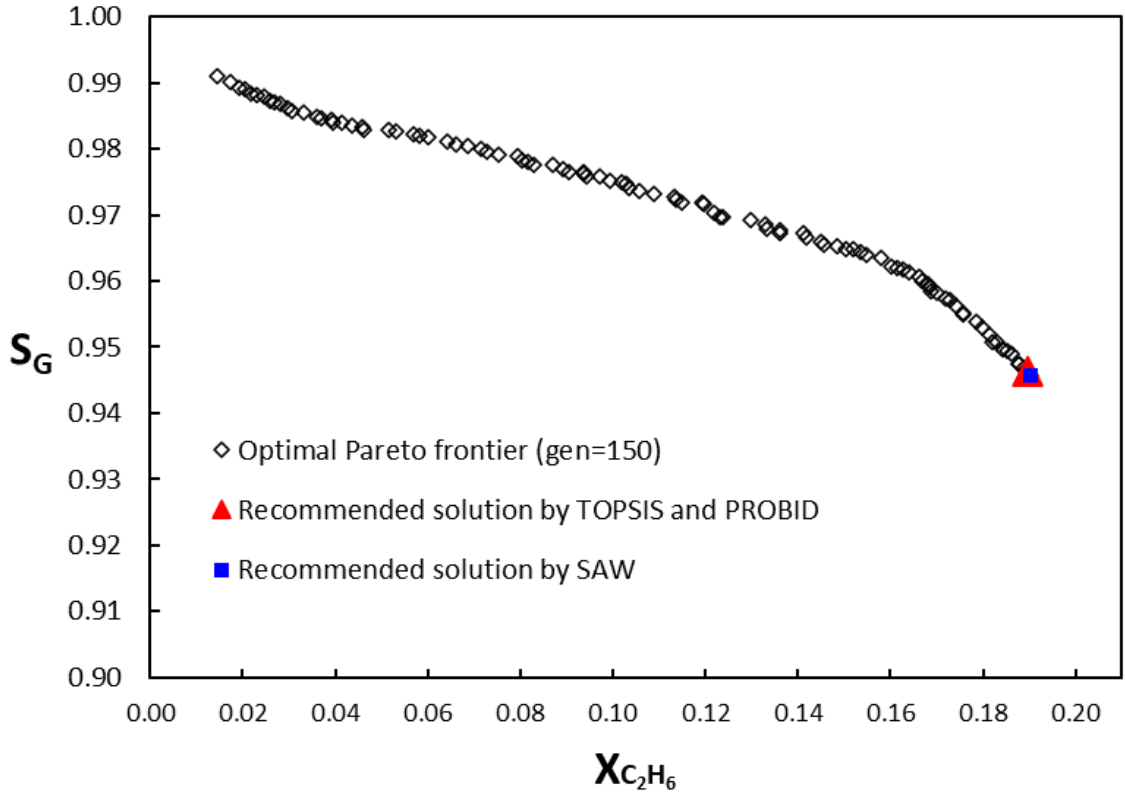

Figure 7. Final Pareto-optimal solutions (◇) at gen=150 for co-maximization of ethane conversion and ethylene selectivity; recommended solution by TOPSIS and PROBID (▲); recommended solution by SAW (■)

Table 5. Values of objectives and decision variables of recommended solutions by TOPSIS, PROBID, and SAW for co-maximization of ethane conversion rate and ethylene selectivity

|  | TOPSIS | PROBID | SAW |
|---|---|---|---|
| $X_{C_2H_6}$ | 0.1896 | 0.1896 | 0.1903 |
| $S_G$ | 0.9464 | 0.9464 | 0.9458 |
| $y_{O_2,0}$ | 0.0185 | 0.0185 | 0.0189 |
| $F_0$ (kmol/h) | 1000 | 1000 | 1000 |
| $T_0$ (°C) | 392.40 | 392.40 | 392.51 |
| $P_{S0}$ (atm) | 5.586 | 5.586 | 5.586 |
| $T_{S0}$ (°C) | 148.01 | 148.01 | 144.23 |
| $F_{S0}$ (kmol/h) | 3539.80 | 3539.80 | 3536.92 |

## 4.3. Maximization of Yearly Ethylene Production Rate and Minimization of Yearly Carbon Dioxide Production Rate

Figure 8 depicts the optimal Pareto frontiers for maximization of yearly ethylene production rate and minimization of yearly carbon dioxide production rate. It includes the initial population alongside the optimal Pareto frontiers at the 25$^{th}$, 100$^{th}$, and 150$^{th}$ generations. A clear correlation emerges between the increase in the carbon dioxide production rate with the ethylene production rate, and vice versa.

As in the MOO scenario above, the initial population size not reach the target of 100 solutions, but widens as generations progress. Examination of the optimal Pareto frontier also reveals broader span of the solutions in the 100$^{th}$ and 150$^{th}$ generations compared to the 25$^{th}$.

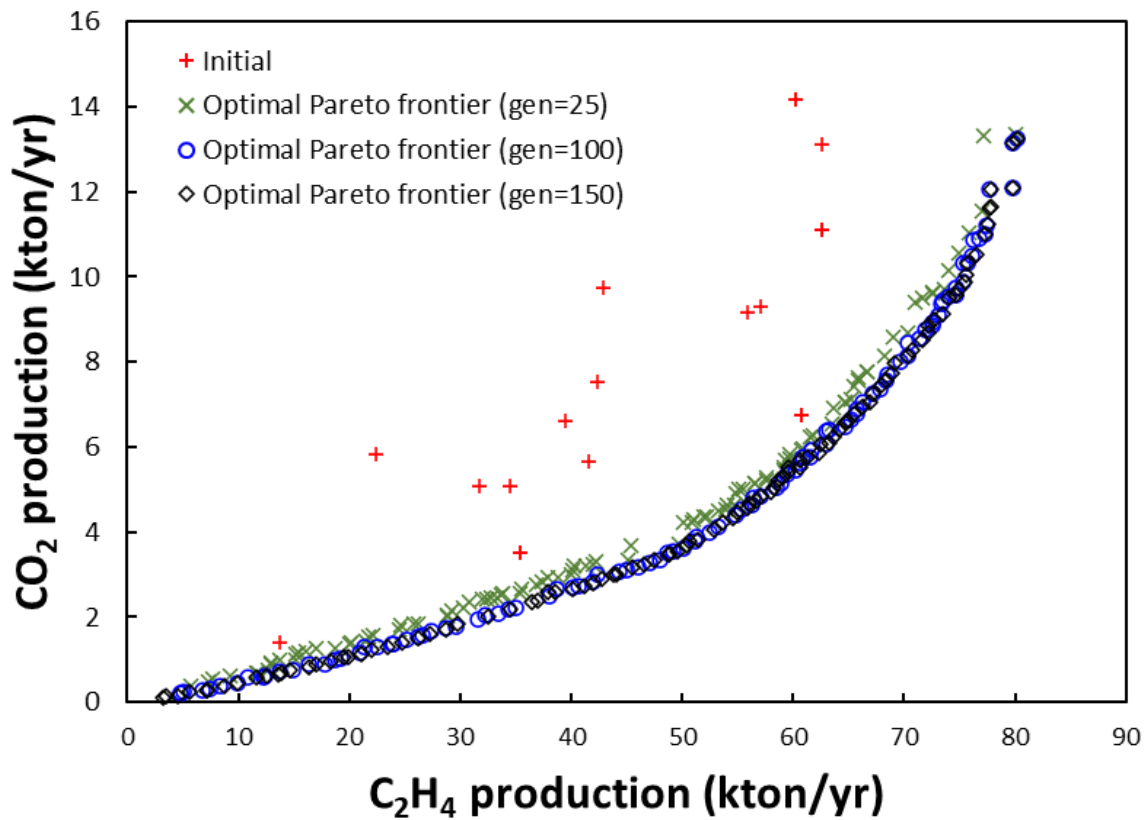

Figure 8. Optimal Pareto frontiers for maximization of yearly ethylene production rate and minimization of yearly carbon dioxide production rate

In addition, a Pareto frontier solution obtained using TOPSIS, PROBID, and SAW methods is selected. In this case, with two objectives, maximizing the yearly ethylene production rate and minimizing the yearly carbon dioxide production rate, equal weights are assigned to both objectives in the MCDM process (i.e., 0.5 to each). TOPSIS and PROBID recommend the same solution (shown as red triangle ▲ in Figure 9). In contrast, the solution recommended by SAW (shown as blue square ■ in Figure 9) diverges markedly from the solution provided by TOPSIS and PROBID. Table 6 presents the values of objective and decision variables for the solutions recommended by TOPSIS, PROBID, and SAW. As can be seen, the solution recommended by SAW is very conservative and environmentally friendly, which is manifested in a remarkably low carbon dioxide production rate of 0.102 kton/yr accompanied by a moderate ethylene production rate of 3.172 kton/yr.

In addition, a Pareto frontier solution obtained using TOPSIS, PROBID and SAW methods is selected. In this case, with two objectives, maximizing the annual rate of ethylene production and minimizing the annual rate of carbon dioxide production, equal weights are assigned to both objectives in the MCDM process (i.e., 0.5 to each). TOPSIS and PROBID recommend the same solution (shown as red triangle ▲ in Figure 9). In contrast, the solution recommended by SAW (shown as blue square ■ in Figure 9) diverges markedly from the solution provided by TOPSIS and PROBID. Table 6 presents the values of the target and decision variables for the solutions recommended by TOPSIS, PROBID and SAW. As can be seen, the solution recommended by SAW is very conservative and environmentally friendly, which is manifested in a remarkably low carbon dioxide production rate of 0.102 kton/year, accompanied by a moderate ethylene production rate of 3.172 kton/year.

On the other hand, the solution suggested by TOPSIS and PROBID present a more balanced approach, that is, a moderate carbon dioxide production rate of 3.457 kton/yr and ethylene production rate of 48.921 kton/yr. It is noticeable that in terms of the decision variables, both the solution selected by SAW and that of TOPSIS and PROBID exhibit similar or equal values of $T_0$, $P_{S0}$, $T_{S0}$, and $F_{S0}$ variables. The starkest contrast arises in the values of $y_{O2,0}$ (i.e., 5.97E-05 for SAW and 1.03E-03 for TOPSIS and PROBID) and $F_0$ (i.e., 1000 by SAW and 4146.31 by TOPSIS and PROBID). Overall, according to the majority vote, the solution selected by TOPSIS and PROBID, along with the corresponding values of the decision variables presented in Table 6, stands as the preferred recommendation for implementation.

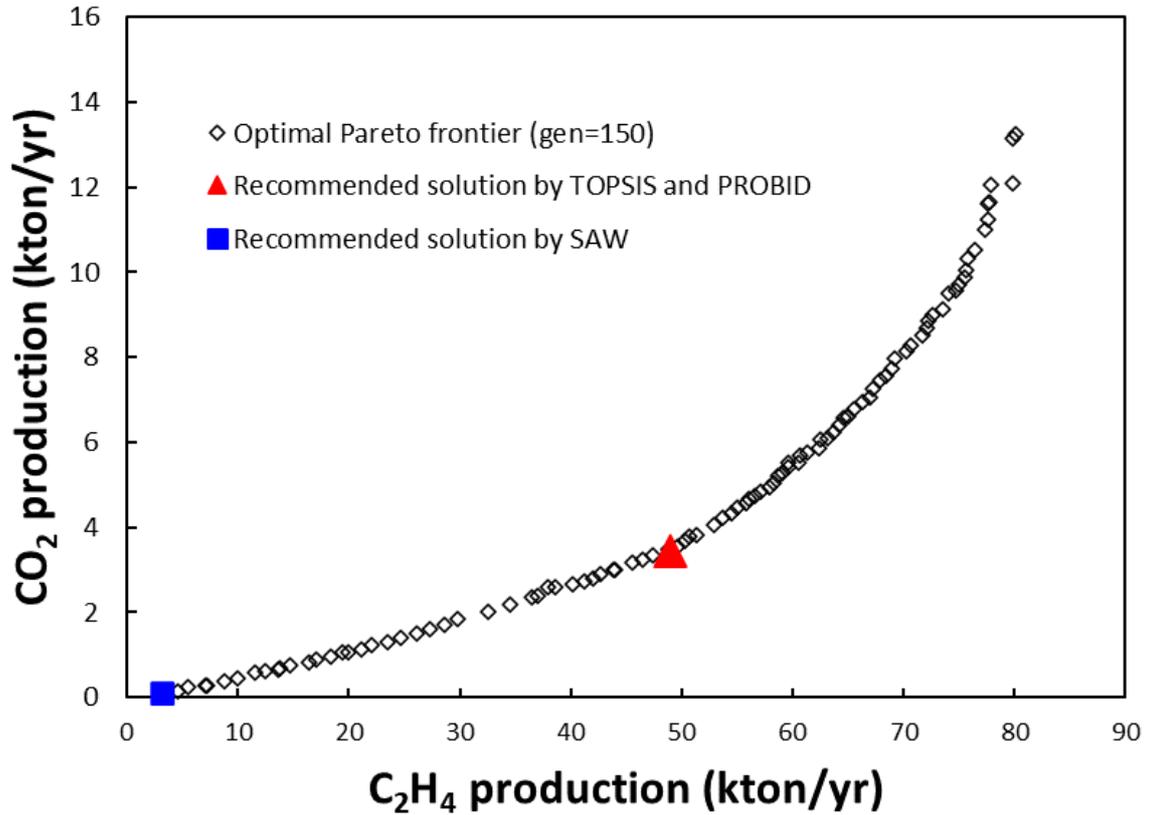

Figure 9. Final Pareto-optimal solutions (◇) at gen=150 for maximization of yearly ethylene production rate and minimization of yearly carbon dioxide production rate; recommended solution by TOPSIS and PROBID (▲); recommended solution by SAW (■)

Table 6. Values of objectives and decision variables of recommended solutions by TOPSIS, PROBID, and SAW for maximization of yearly ethylene production rate and minimization of yearly carbon dioxide production rate

|  | TOPSIS | PROBID | SAW |
|---|---|---|---|
| $C_2H_4$ production (kton/yr) | 48.921 | 48.921 | 3.172 |
| $CO_2$ production (kton/yr) | 3.457 | 3.457 | 0.102 |
| $y_{O2,0}$ | 1.03E-03 | 1.03E-03 | 5.97E-05 |
| $F_0$ (kmol/h) | 4146.31 | 4146.31 | 1000 |
| $T_0$ (°C) | 420 | 420 | 420 |
| $P_{S0}$ (atm) | 5.481 | 5.481 | 5.499 |
| $T_{S0}$ (°C) | 196.18 | 196.18 | 200 |
| $F_{S0}$ (kmol/h) | 4049.68 | 4049.68 | 4173.09367 |

## 4.4. Simultaneous Maximization of Ethane Conversion and Ethylene Production Rate and Minimization of Carbon Dioxide Production Rate

The previous prior two subsections elucidated the competing relationships between ethane conversion rate maximization and ethylene selectivity maximization, as well as between ethylene production rate maximization and carbon dioxide production rate minimization. In this subsection, a three-objective optimization formulation that pursues simultaneous maximization of ethane conversion and ethylene production rate and minimization of carbon dioxide production rate is presented. The 3D plot in Figure 10 shows the final optimal Pareto frontier at the 150th generation of NSGA-II.

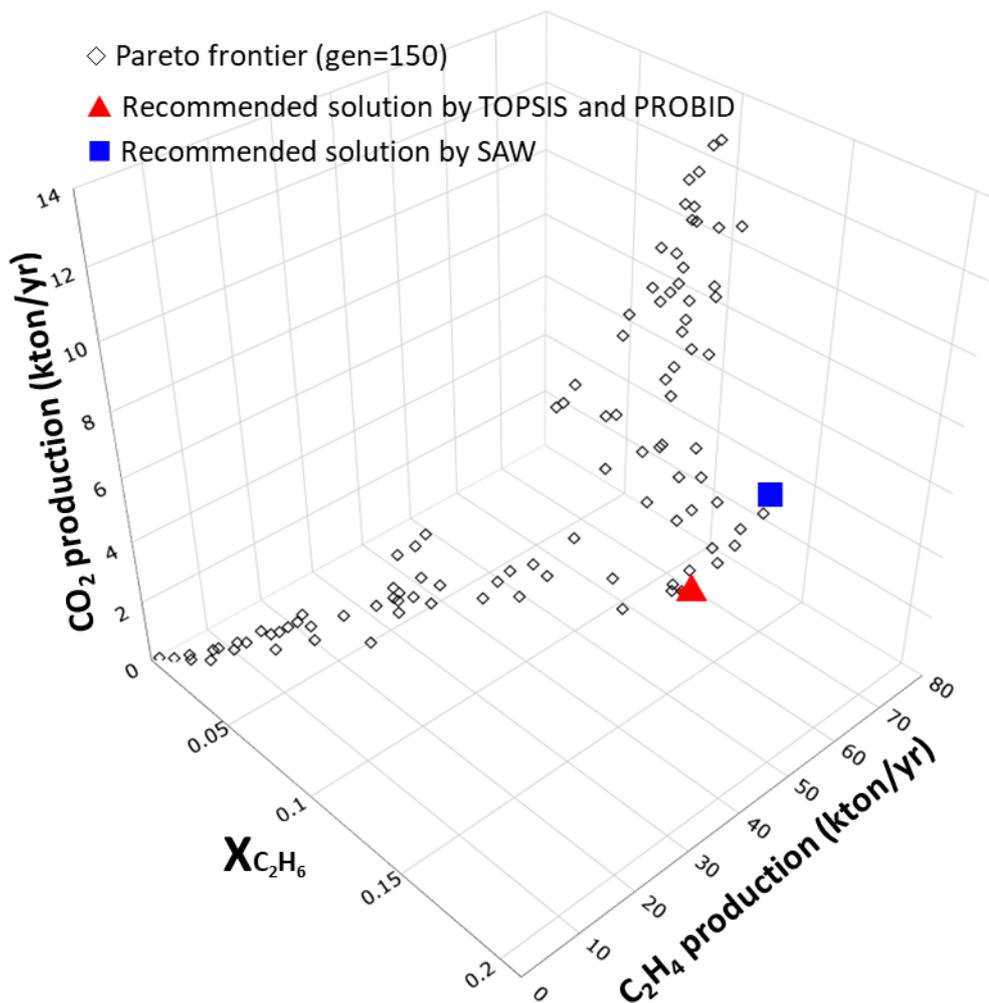

Figure 10. Final Pareto-optimal solutions (◇) at gen=150 for simultaneous maximization of ethane conversion and ethylene production rate and minimization of carbon dioxide production rate; recommended solution by TOPSIS and PROBID (▲); recommended solution by SAW (■)

Subsequently, TOPSIS, PROBID, and SAW methods are used to rank the solutions on the final optimal Pareto frontier and to identify a recommended solution. In this scenario, where the three objectives are the simultaneous maximization of ethane conversion and ethylene production rate, and the minimization of carbon dioxide production rate, each objective receives equal weight in the MCDM process (i.e., 1/3 to each). Both TOPSIS and PROBID choose the same solution (shown as red triangle ▲ in Figure 10). On the other hand, the solution recommended by SAW (shown as blue square ■ in Figure 10) shows some deviation, although not drastic. Table 7 presents the values of objectives and decision variables for the solutions recommended by TOPSIS, PROBID, and SAW. From the analysis, SAW recommends an ethane conversion rate of 0.2049, which is slightly above the 0.1810 recommended by TOPSIS and PROBID. Similarly, the ethylene production rate suggested by SAW is 42.123 kton/yr, higher than the 38.196 kton/yr predicted by TOPSIS and PROBID. Consequently, SAW predicts a higher carbon dioxide production rate of 9.304 kton/yr, compared to 6.011 kton/yr for TOPSIS and PROBID.

Additionally, in terms of decision variables, both the solution selected by SAW and the one chosen by TOPSIS and PROBID coincide in a $F_0$ value of 1000 kmol/h. However, moderate differences are observed in the remaining five decision variables ($y_{O2,0}$, $T_0$, $P_{S0}$, $T_{S0}$, and $F_{S0}$). Overall, based on majority voting and similarity to that chosen by SAW, the solution proposed by TOPSIS and PROBID, along with the values of the associated decision variable detailed in Table 7, stands as the preferred recommendation for implementation.

Table 7. Values of objective and decision variables of recommended solutions by TOPSIS, PROBID, and SAW for simultaneous maximization of ethane conversion and ethylene production rate and minimization of carbon dioxide production rate

|  | TOPSIS | PROBID | SAW |
|---|---|---|---|
| $X_{C_2H_6}$ | 0.1810 | 0.1810 | 0.2049 |
| $C_2H_4$ | 38.196 | 38.196 | 42.123 |
| $CO_2$ | 6.011 | 6.011 | 9.304 |
| $y_{o2,0}$ | 0.0108 | 0.0108 | 0.0180 |
| $F_0$ | 1000 | 1000 | 1000 |
| $T_0$ | 403.77 | 403.77 | 394.98 |
| $P_{S0}$ | 5.794 | 5.794 | 5.839 |
| $T_{S0}$ | 188.48 | 188.48 | 128.31 |
| $F_{S0}$ | 4076.60 | 4076.60 | 4101.79 |

## 5. Conclusions

In conclusion, this work delved deeply into the search and detection of the optimal operating conditions for the ODHE process carried out in a packed-bed multi-tubular membrane reactor with oxygen permeation through the membrane tubes, simultaneously considering multiple conflicting objectives inherent in such a system. To overcome the complex optimization challenges, the MOO method based on genetic algorithm, NSGA-II is integrated with three MCDM methods: TOPSIS, PROBID, and SAW. Then, three distinct MOO scenarios are rigorously examined: joint the maximization of ethane conversion rate and ethylene selectivity; the maximization of yearly ethylene production rate and minimization of yearly carbon dioxide production rate; and the simultaneous maximization of ethane conversion and ethylene production rate and minimization of carbon dioxide production rate. By harnessing the power of NSGA-II, the optimal Pareto frontiers are obtained for these scenarios, followed by the application of the three MCDM methods mentioned above to rank the non-dominated solutions, recommending a single solution for final application in each scenario. The approach adopted in this work provides detailed insights into the trade-offs among the multiple conflicting objectives, improves understanding of their relationships, and reveals the impacts of the decision variables on the objective variables, thus enabling practitioners to make judicious and informed decisions in the ODHE process operation.